\documentclass[conference]{IEEEtran}
\IEEEoverridecommandlockouts
\usepackage[utf8]{inputenc}
\usepackage{cite}
\usepackage{amsmath,amssymb,amsfonts}
\usepackage{bm}
\usepackage{cancel}
\usepackage[ruled,linesnumbered]{algorithm2e}
\usepackage{algorithmic}
\usepackage{graphicx}
\usepackage{textcomp}
\usepackage{xcolor}
\usepackage[export]{adjustbox}
\usepackage{xurl}
\usepackage{comment}
\usepackage{paralist} 
\usepackage{enumitem} 
\usepackage[T1]{fontenc}
\DeclareMathAlphabet{\mathpzc}{OT1}{pzc}{m}{it}
\usepackage{tabularx}
\usepackage{booktabs}
\usepackage{multirow}
\usepackage{makecell}
\usepackage{soul}
\usepackage{subfigure}
\usepackage{balance}
\usepackage{hhline}
\usepackage{booktabs}
\usepackage{enumitem}
\usepackage[acronym]{glossaries}

\usepackage{comment}

\usepackage[most]{tcolorbox}

\usepackage{nomencl}
\makenomenclature
\usepackage{etoolbox}
\renewcommand\nomgroup[1]{%
  \ifstrequal{#1}{P}{\vspace{10pt}\item[\textbf{Parameters}]}{%
  \ifstrequal{#1}{V}{\vspace{10pt}\item[\textbf{Variables}]}{}}{%
  \ifstrequal{#1}{S}{\vspace{10pt}\item[\textbf{Sets}]}{}}%
}

\title{Sensitivity Analyses of Resilience-oriented Risk-averse Active Distribution Systems Planning}

\author{\IEEEauthorblockN{Abodh Poudyal and Anamika Dubey}
\IEEEauthorblockA{Washington State University\\
Pullman, Washington, USA\\
Email: \{abodh.poudyal, anamika.dubey\}@wsu.edu
\vspace{-3em}}
\thanks{This work was supported by National Science Foundation (NSF) CAREER Grant \#1944142.}
}

\begin{document}
\bstctlcite{IEEEexample:BSTcontrol}
\maketitle
\thispagestyle{plain}
\pagestyle{plain}

\begin{abstract}
This paper presents sensitivity analyses of resilience-based active distribution system planning solutions with respect to different parameters. The distribution system planning problem is formulated as a two-stage risk-averse stochastic optimization model with conditional value-at-risk (CVaR) as the risk measure. The probabilistic scenarios are obtained using regional wind profiles, and Monte Carlo simulations are conducted to obtain failure scenarios based on component fragility models. The planning measure includes advanced distribution grid operations with intentional islanding measures. The three main parameters used in this work for sensitivity analysis are the number of scenarios, risk preference, and planning budget allocation.
Such analysis can provide additional information to system operators on dispatching the planning budget and available resources properly to enhance the grid's resilience.
\end{abstract}


\vspace{-0.7em}
\section{Introduction}
Extreme weather-related events, often termed as high-impact, low-probability (HILP) events, have created devastating impacts on critical infrastructures, especially electric power systems. The power distribution systems are even more affected as they cannot withstand such extreme weather events and are more vulnerable than the bulk electric grid. It is reported that there has been a 67\% increase in power outages due to weather-related events since 2000, with 80\% -90\% of outages resulting from failures in power distribution systems~\cite{noaa, DOE_res}. Furthermore, the frequency of these extreme events has increased at an alarming rate, so it is important to make the power distribution system resilient against these extreme events.

Existing works generally model power distribution resilience planning problems as stochastic programming or robust optimization models. In~\cite{7381672}, a robust optimization framework is proposed for resilience distribution systems planning against extreme weather events. Line hardening and backup distributed generation (DG) resources are used as the planning decisions to enhance the grid's resilience.  Similarly,~\cite{8329529} proposes a resilience-oriented two-stage stochastic optimization model with multiple planning decisions. Another paper proposed a proactive operational planning measure for restoring damaged components using repair crews~\cite{8409997}. However, the planning is based on the objective of minimizing the expected value of second-stage cost, and each of the scenarios is considered to have an equal probability of occurrence. While such methods are popular in the literature, they are unsuitable for resilience-oriented planning focusing on HILP events. Thus, HILP events should be explicitly modeled in the objective. Several efforts have been to define risk metrics for resilience in power systems. One commonly used risk-based metric is Conditional value-at-risk (CVaR)~\cite{9810633, poudel2019risk}. Previously, CVaR, a risk metric, has been used in distribution systems planning~\cite{khodabakhsh2016optimal, Wu2021, 9942328}.       
Although multiple resilience planning frameworks have been proposed, a majority of the works do not analyze the sensitivity of various parameters that can affect the solutions of the resilience-oriented planning problem in the power distribution grid. Some works address the computational complexity of the number of scenarios in the stochastic optimization model but fail to show if the quality of planning decisions changes based on the number of scenarios. Resilience-based investments are very expensive from the system operator's point of view as such investments are made for HILP events, which occur occasionally but create significant damage. Hence, it is necessary to understand the decisions made from every possible angle to justify the current investments for situations when the scenarios are realized. The goal of this paper is to evaluate how optimal planning decisions vary based on multiple systems and optimization parameters. The selected parameters include the total budget allocated, the number of candidate locations to install DGs, and the risk preference of the system operator. We employ our previously proposed two-stage risk-averse stochastic optimization framework for grid resilience planning for sensitivity analysis.        


\vspace{-0.75em}
\section{Distribution Systems Planning Model}\label{sec:planning_model}
\begin{figure}[t]
    \centering
    \includegraphics[width=\linewidth]{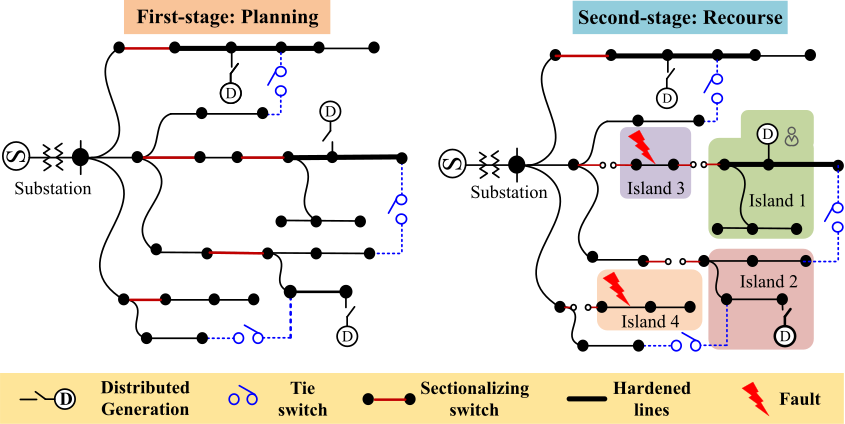}
    \caption{Two-stage planning framework example for a specific scenario.}
    \vspace{-0.6cm}
    \label{fig:model_representation}
\end{figure}

The distribution system planning model is represented as a generic two-stage stochastic optimization problem, where the first-stage decisions guide the planning cost, and the second-stage decisions are the scenario-specific operational decisions. To include the impacts of HILP events and identify the trade-off between risk-neutral and risk-averse planning decisions, a two-stage risk-averse planning model is used~\cite{9942328}. The planning decisions are DG locations and sizes, whereas the second stage provides the set of operational decisions to minimize the prioritized load loss once an event is realized. The event is modeled via regional wind profile, and line fragility models are used to assess the impact of wind storm events on the distribution system. As discussed in~\cite{7801854}, it is assumed that the uncertain wind storm event has some form of a probability distribution. Once an event is realized, the planned DGs, with grid-forming capabilities, can form intentional islands to energize the disconnected part of the grid. Fig.~\ref{fig:model_representation} shows the general framework of the two-stage planning model for a specific scenario. When multiple sections of a distribution system are disconnected from the main feeder, the sections with DGs are energized, as shown in the figure. The planning decisions made in the first stage should remain optimal for every scenario realized in the second stage. Please refer to ~\cite{9942328} for details on the overall simulation framework. 

The solution of stochastic optimization remains optimal for the selected set of scenarios and other assumptions in the overall framework~\cite{asi2019importance}. It is important to ensure that the selected scenarios represent several other scenarios and that the assumptions are valid. Hence, several parameters can affect the planning solution. For instance, assigning more budget on the DG-based resources can help pick up more loads once an event is realized. However, the amount of load picked up can be saturated at some point when the intensity of the event is extreme such that it is impossible to pick up any load. At that point, investing more in DG-based resources would not make much sense. Thus, the remainder of the paper identifies some important parameters and the sensitivity of the optimization solution with those parameters. 

\vspace{-0.5em}
\section{Methodology}\label{sec:methodology}
\subsection{Mathematical Formulation}
The distribution systems planning model can be represented as a two-stage risk-averse stochastic planning model, as shown in (\ref{eq:risk_twostage}):

\begin{equation}
    \small
    \min~(1-\lambda)\mathbb{E}[\mathcal{Q}(\delta,\mathcal{E})] + \lambda CVaR_\alpha(\mathcal{Q}(\delta,\mathcal{E}))\\
    \label{eq:risk_twostage}
\end{equation}
\noindent
where 
\begin{equation*}
\small
    \begin{gathered}
        \mathbb{E}(Q(\delta, \mathcal{E})) := \left(\sum_{\xi \in \mathcal{E}}\sum_{i\in  \mathcal{B}_S}\sum_{\phi\in\{a,b,c\}}(1 - s^\xi_i)\ w_i\ P_{Li}^{\phi,\xi}\right)\\
        CVaR_\alpha(Q(\delta, \mathcal{E})) := \left(\eta + \frac{1}{1-\alpha}\sum_{\xi \in \mathcal{E}}p^\xi\nu^\xi\right)
    \end{gathered}
\end{equation*}
\noindent

\noindent
where $\mathbb{E}[\mathcal{Q}(\delta,\mathcal{E})]$ is the expected value of the prioritized load loss in the second stage for all scenarios $\xi \in \mathcal{E}$ and first stage decision $\delta$ that decides the location and size of DG, $\mathcal{B}_S$ is the set of energized buses, $\phi$ represent the set of three phases, $s_i \in {0,1}$ represents the load pick-up status for loads in each bus $i$ at scenario $\xi$, $w_i$ is the weight defining priority of load connected to bus $i$, and $P_{Li}^{\phi,\xi}$ represents the three-phase active power load at bus $i$ for scenario $\xi$. $CVaR_\alpha(\mathcal{Q}(\delta,\mathcal{E}))$ is the $CVaR$ of the second stage at confidence level $\alpha$, $p^\xi$ is the probability of scenario $\xi \in \mathcal{E}$, $\nu^\xi$ is the excess $CVaR$ parameter for scenario $\xi$ such that $\nu^\xi \geq \mathcal{Q}(x,\xi) - \eta$, and $\lambda$ is the risk trade-off parameter. Here, $\eta \in \mathbb{R}$ is value-at-risk (VAR) and is the first-stage variable.

The first-stage budget constraint for (\ref{eq:risk_twostage}) is given by (\ref{eq:DG_size}). It ensures that the budget for DG installation (siting and sizing) should be less than $C_{max}^{DG}$. $\delta^{DG}_i \in {0,1}$ and $\beta_i^{DG} \in \mathbb{R}^+$ are the first-stage decision variables deciding the location and size of DG, respectively. It is assumed that the per unit cost of DG sizing and siting are the same for all candidate nodes. Hence, the total size of the DG to be installed decides the overall budget. These assumptions can easily be relaxed by using actual costs for each resource, if available. 

\vspace{-0.5em}
\begin{equation}
\small
    \sum_{i\in \mathcal{B}_{DG}} c_i^{DG}\delta_i^{DG} \beta^{DG}_i \leq \mathcal{C}^{DG}_{max} \\
    \label{eq:DG_size}
\end{equation}

This work does not describe the second-stage scenario-dependent restoration model for the lack of space and is discussed in detail in~\cite{9942328}. The second stage of the framework comes into play once a scenario is realized. However, the first-stage decision should remain optimal for each scenario realization of the second stage.

\begin{figure}[!t]
     \centering
     \subfigure[]
     {
        \includegraphics[width=0.57\linewidth]{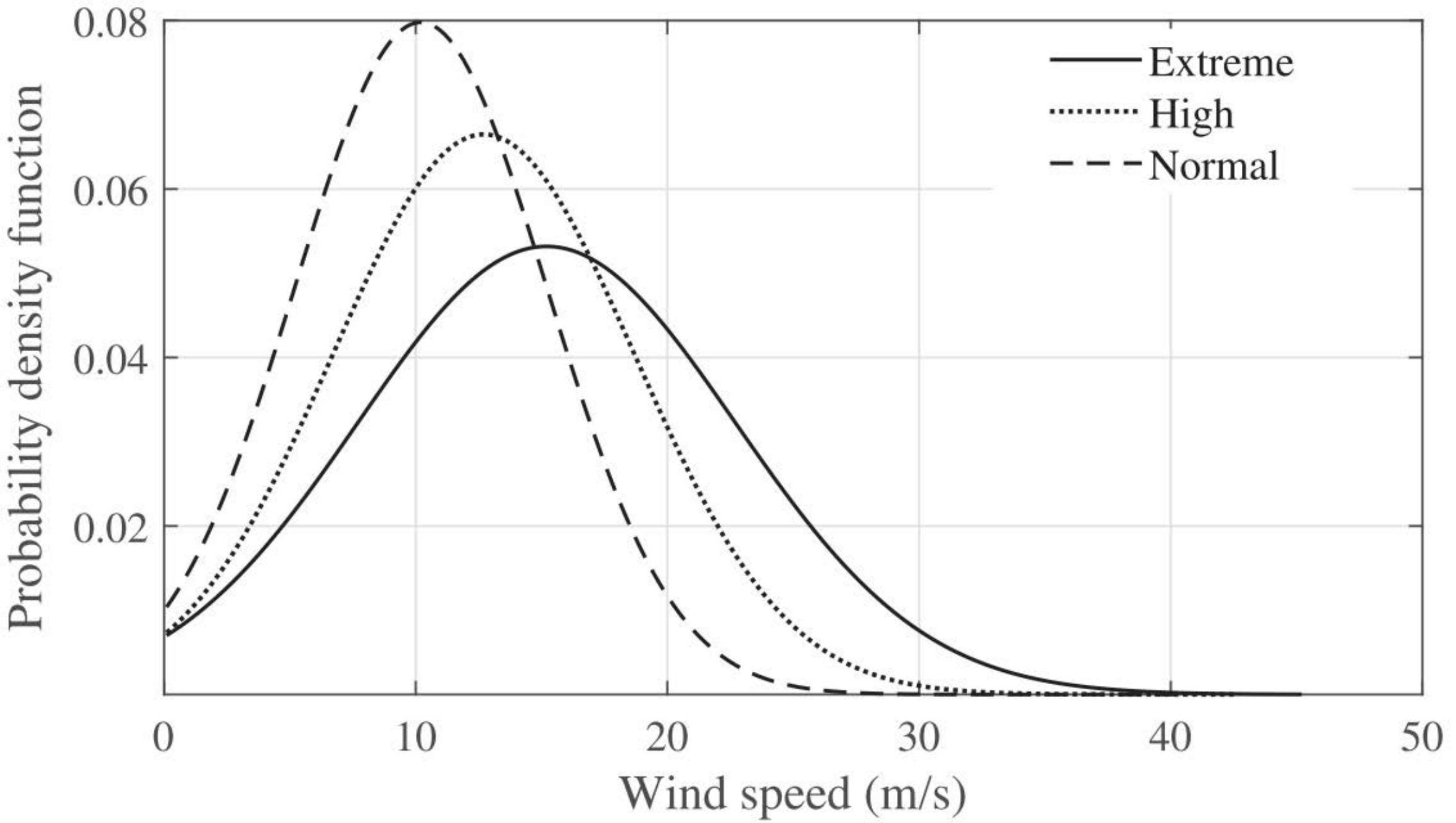}
        \label{fig:wind_profile}
    }\vspace{-0.1cm}
    \subfigure[]
    {
        \includegraphics[width=0.36\linewidth]{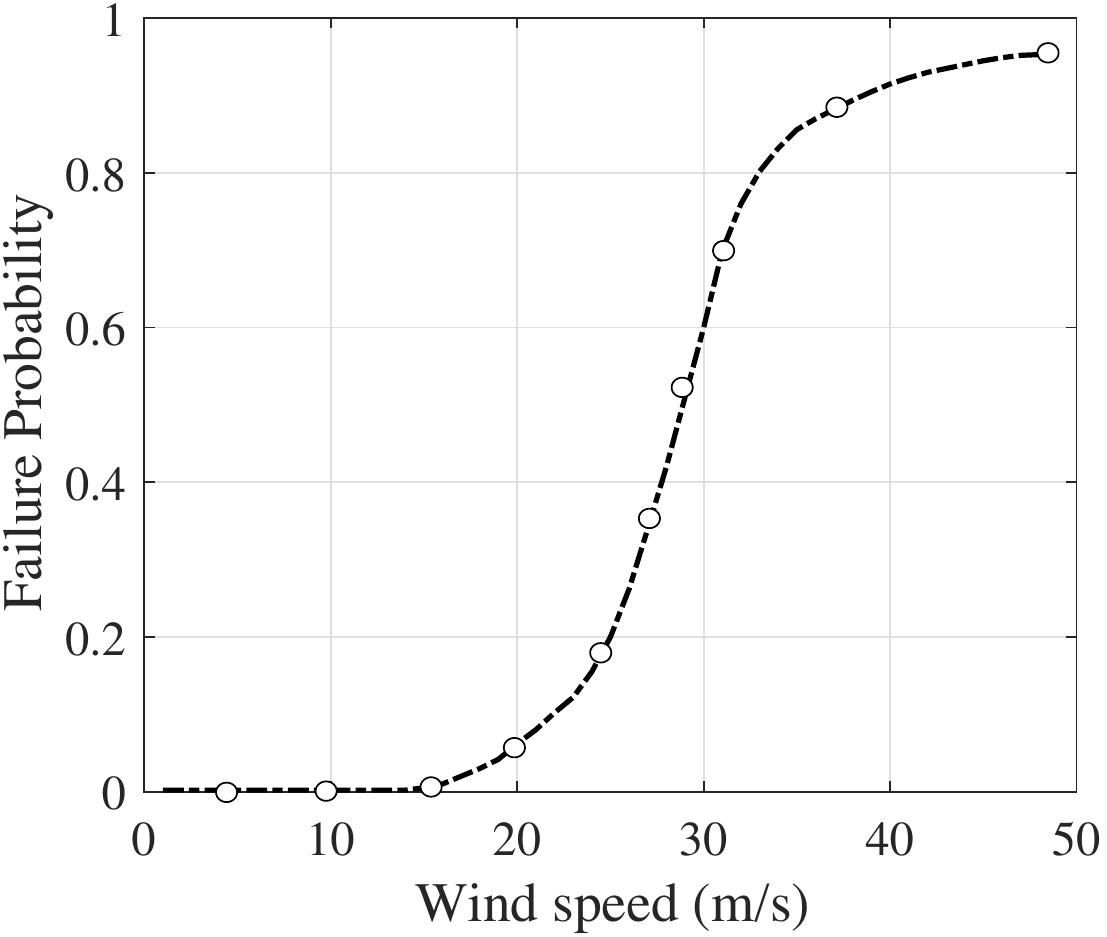}
        \label{fig:fragility}
     }
     \vspace{-0.3cm}
     \caption{a) Regional wind profile of varying intensity~\cite{poudel2019risk}. b) Line Fragility curve with respect to wind speed~\cite{panteli2017metrics}.}
     \label{fig:wind_fragility}
     \vspace{-0.5cm}
\end{figure}

\begin{figure*}[!t]
        \centering
        \includegraphics[width=\linewidth]{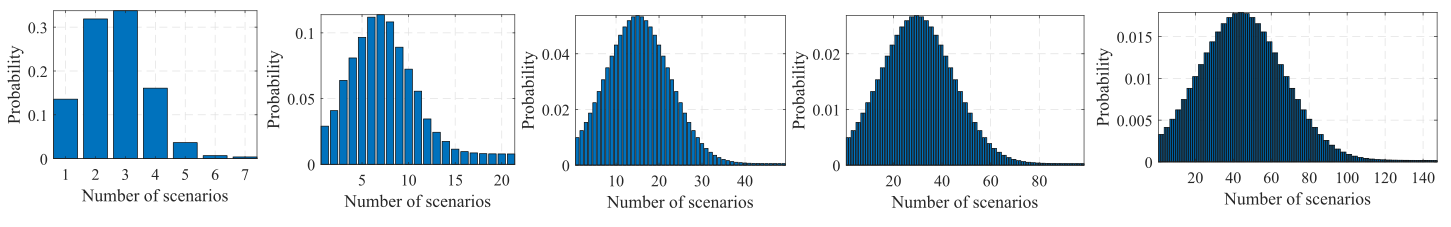}
        \vspace{-2.5em}
        \caption{Representative number of scenarios with their respective probabilities: (a) $n = 7$, (b) $n = 21$, (c) $n = 49$, (d) $n = 98$ e) $n = 147$.}
        \label{fig:scen_probab}
        \vspace{-1.2em}
\end{figure*}

\vspace{-0.75em}
\subsection{Simulation Framework}
The regional wind profile shown in Fig.~\ref{fig:wind_profile} is used to generate wind storm scenarios. The distribution system is relatively small compared to the bulk grid; hence, it can be assumed that the entire system experiences wind storms of similar intensity~\cite{9917119}. Similarly, Fig.~\ref{fig:fragility} shows the fragility of a distribution line to the wind speed experienced by the same. Fragility models have been widely adopted for impact assessment of power system components subject to weather-related events. Since fragility models provide a probabilistic assessment of failures, Monte Carlo simulations (MCS) are conducted to observe failure scenarios. Based on the converged value of MCS, a smart scenario selection strategy is used to select representative scenarios for several cases, as discussed in~\cite{9942328}. The simulations are conducted for different parameters in the two-stage framework to understand the variation of planning solutions with respect to parameter changes.

\section{Planning Model Sensitivity Analysis}\label{sec:sensitivity_results}
\begin{figure}[t]
    \centering
    \includegraphics[width=0.8\linewidth]{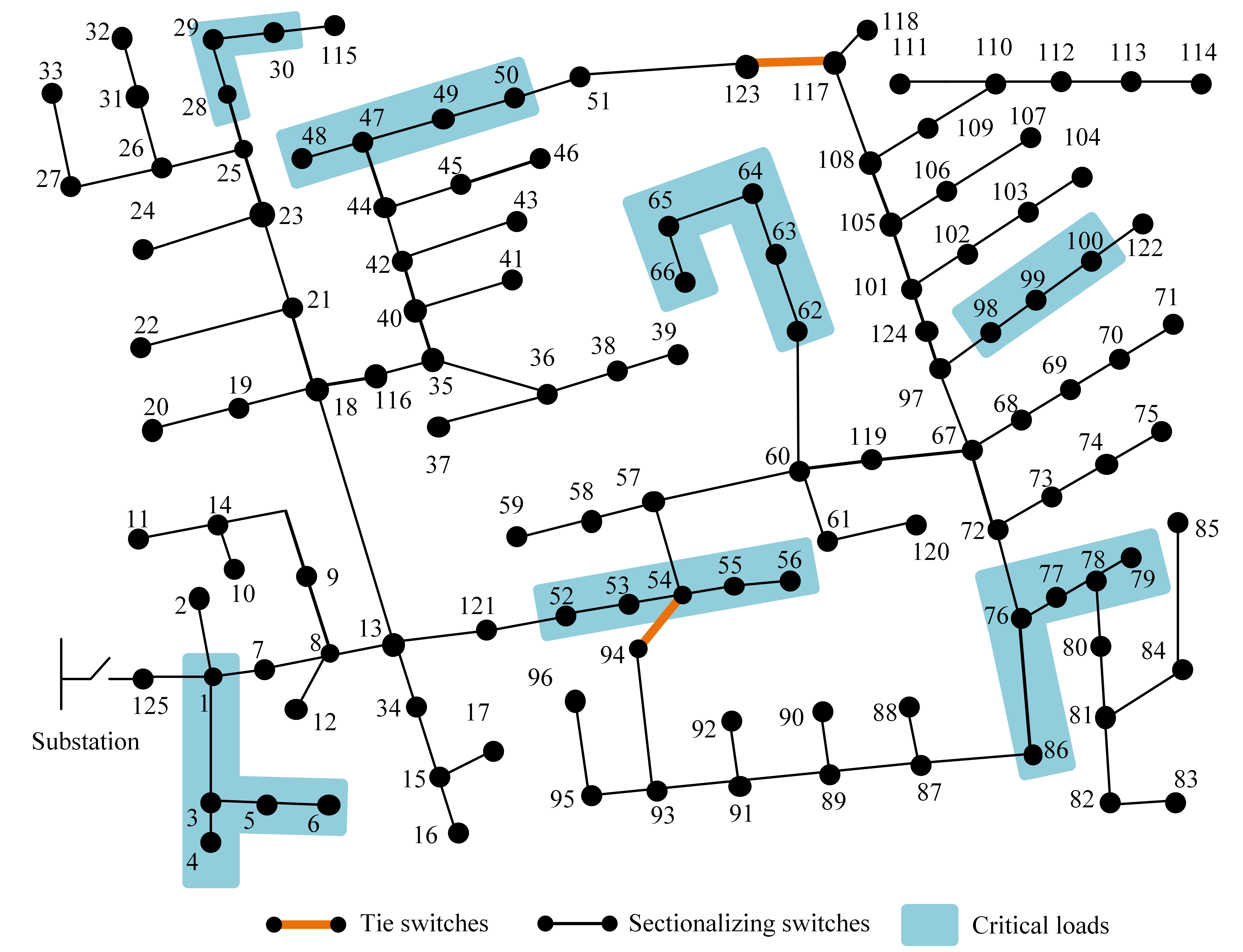}
    \caption{Modified IEEE-123 bus system}
    \label{fig:123bus}
    \vspace{-1.5em}
\end{figure}

Although the two-stage planning framework provides an optimal planning decision for several scenarios, the decisions can be sensitive to assumptions and considerations made during the planning or operational stage. This section discusses the sensitivity of the two-stage planning model for different parameters. In this work, we focus on the selection of the number of scenarios, risk preference and confidence level, initial planning budget, and selection of candidate DG locations. The simulations are conducted on a modified IEEE-123 bus system with several sectionalizing and tie-line switches, see Fig.~\ref{fig:123bus}. The total non-prioritized active power demand of the system is $4485$ kW whereas the prioritized active power demand, $\sum_i{w_i P_{Li}}$, is $20775$ kW for $w_i=10$. In this work, prioritized load loss is used for analysis. The risk-averse two-stage optimization model is formulated using the PySP module in Pyomo~\cite{watson2012pysp} and solved as a large mixed-integer linear programming (MILP) problem using commercial off-the-shelf solvers. 

\vspace{-0.5em}
\subsection{Number of Scenarios}
\begin{figure}[t]
    \centering
    \includegraphics[width=0.85\linewidth]{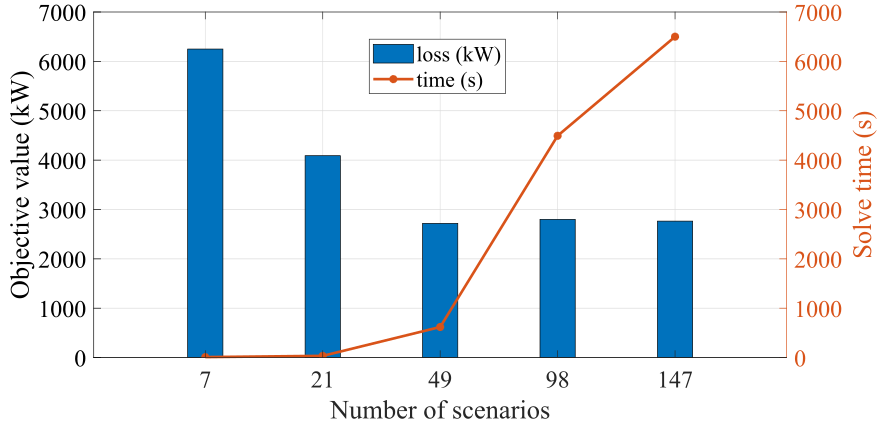}
    \caption{Variation of objective value and solve time of a planning problem with respect to the number of scenarios.}
    \label{fig:scen_vs_time}
    \vspace{-1.5em}
\end{figure}

Let $\mathcal{E} = \{\xi_1, \xi_2, ..., \xi_n\}$ be the set of scenarios under consideration such that each set $\mathcal{E}$ contains $n$ number of scenarios. Fig.~\ref{fig:scen_probab} shows the scenarios and their respective probabilities of occurrence for five different values of $n$. Similarly, the changes in the objective value with respect to the selected number of scenarios and solve times are shown in Fig.~\ref{fig:scen_vs_time}. For simplicity, only the expected value of the overall objective is observed here. However, the analysis remains true for any other objectives. For each $n$, the result is observed by taking an expected value over ten sets of scenarios. For the minimization problem, the objective value decreases from $n = 7$ to $n = 21$ with minimal change in the solve time. The results for $n = 98$ and $n = 147$ (2764.16 kW) are rather similar to the ones obtained for $n = 49$ (2719.17 kW). However, the solve time for the former two cases is significantly higher than the one with $n = 49$. Hence, it is clear that for $n = 49$, the planning decisions are acceptable considering both the perspective of loss minimization and solve time. Although it is desired to include all of the scenarios in the optimization model, it is computationally inexpensive compared to the improvement in solution quality. Hence, using too few scenarios in a stochastic optimization model would be faster to solve but at the expense of poor solution quality, and the contrary is true when using a higher number of scenarios.  

\vspace{-0.25em}
\subsection{Risk preference}
\begin{figure}[t]
    \centering
    \includegraphics[width=0.8\linewidth]{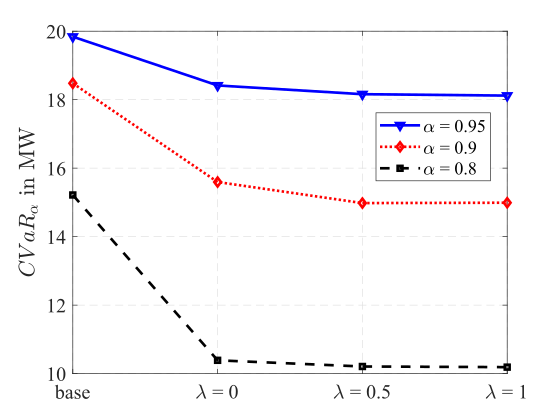}
    \caption{Sensitivity of $CVaR$ of the prioritized load loss with respect to $\alpha$ and $\lambda$.}
    \label{fig:cvar_alpha}
    \vspace{-2em}
\end{figure}

The risk-averse two-stage stochastic optimization problem defined in (\ref{eq:risk_twostage}) is affected by two different risk parameters, $\alpha$ and $\lambda$. The parameter $\lambda$ defines the risk preference of a system operator and is monotonic with the risk aversion, i.e., $\lambda=0$ represents risk neutrality, and the overall objective in  (\ref{eq:risk_twostage}) becomes a general two-stage stochastic optimization problem whereas $\lambda=1$ represents risk aversion. Hence, higher values of lambda focus on risk minimization. Furthermore, $\alpha$ determines the confidence level in $CVaR_\alpha$ and defines the number of scenarios to consider when defining the system risk. For instance, $\alpha = 0.95$ means that only 5\% of the tail event scenarios are risky, and the risk-averse policies are defined for those representative events. However, for lower alpha values, the value of $VAR_\alpha$ decreases, and more scenarios are considered risky as $CVaR_\alpha$ represents the expected value of losses beyond $VaR_\alpha$. This can increase the planning expenses as the system operators have to plan for more events considered as ``risk'' to the system. Fig.~\ref{fig:cvar_alpha} shows the $CVaR_\alpha$ of prioritized load loss for different values of $\alpha$ and $\lambda$. As discussed above, with higher values of $\alpha$, the $CVaR_\alpha$ decreases as the $VAR_\alpha$ decreases, which shifts the $CVaR_\alpha$ to the lower side. Similarly, the change is consistent with higher values of $\lambda$ as the risk preference of the system operator is towards risk minimization. Thus, it is essential to identify appropriate values of $\alpha$ and $\lambda$ to include the risk preference of the system operator.  

\vspace{-0.6em}
\subsection{Budget Allocation and Candidate Locations of DGs}
\begin{figure}[t]
     \centering
     \subfigure[]
     {
        \includegraphics[width=0.8\linewidth]{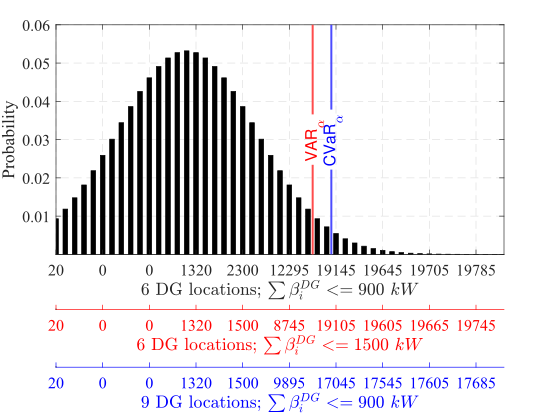}
        \label{fig:DG_neutral}
    }\vspace{-0.1cm}
    \subfigure[]
    {
        \includegraphics[trim ={0 0 0 0.5cm} ,clip, width=0.8\linewidth]{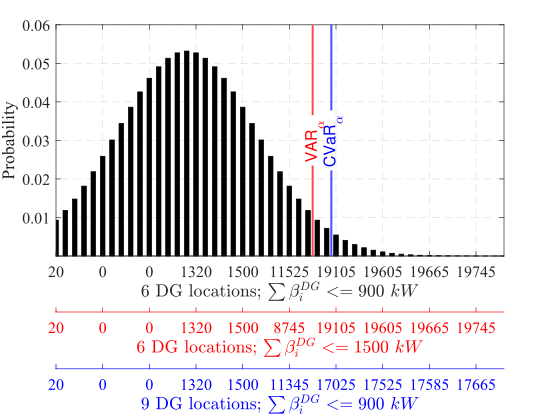}
        \label{fig:DG_averse}
     }
     \vspace{-0.3cm}
     \caption{$CVaR_\alpha$ for different DG budget allocation strategies for a) risk-neutral and b) risk-averse planning. The X-axis represents the prioritized load loss corresponding to each scenario and the y-axis corresponds to $p_\xi$.}
     \label{fig:DG_tradeoffs}
     \vspace{-2em}
\end{figure}

The solution of the two-stage planning model is highly dependent on the overall budget allocation. Here, we control the parameter $C_{max}^{DG}$ as described in (\ref{eq:DG_size}). This refers to the overall size of the DG in kW, i.e., $\sum \beta_i^{DG}$. Furthermore, the solution sensitivity is also observed for the candidate DG locations. Although it is desirable to place DGs in every location and install the size of DG equal to the overall demand of the system, it is often not possible due to the limited budget of the system operator. Hence, it is generally practical to select candidate locations for DGs and limit the overall size, which is guided by the budget of the system operator. Fig.~\ref{fig:DG_tradeoffs} represents the prioritized load loss distribution for a varying number of candidate DG locations, budget allocation, and risk preference. The number of scenarios for each of the cases is 49.

The six candidate locations for DG placement are 95, 122, 39, 85, 56, and 66. Locations 47, 27, and 114 are identified as three additional locations when nine candidate nodes are desired. It is important to note that the actual DG locations are decided by the optimization problem such that $\delta_i^{DG} = 1$ \textit{iff} $\beta_i^{DG} > 0$. Furthermore, $\sum \beta_i^{DG} \leq 900$ kW ensures that the system operators only have a budget to install DGs with a total capacity of $900$ kW.

Interestingly, for both the risk-neutral and risk-averse cases, with a number of candidate locations intact, increasing the overall budget to increase the capacity of DG has minimal effect on the $CVaR_\alpha$. However, the expected value of the prioritized load loss decreases from 2467.46 kW to 2222.43 kW. However, with the overall budget kept intact, increasing additional candidate DG locations decreases the $CVaR_\alpha$ significantly. The effect is even more prominent for risk-averse objective with $CVaR_\alpha$ decreasing to $15811.59$ kW with nine candidate locations as compared to $18119.1$ kW with 6 candidate locations when $\sum \beta_i^{DG} \leq 900$ kW. This leads to an important conclusion that multiple DG locations with small sizes DGs are preferable over a few DG locations with higher sizes DGs when the overall budget is limited.      

\begin{figure}[ht]
    \centering
    \includegraphics[width=0.8\linewidth]{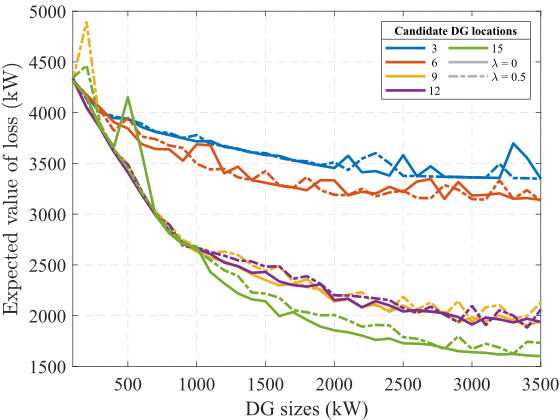}
    \caption{Variation of the expected value of prioritized load loss with respect to the number of candidate DG locations, budget allocation, and risk preference.}
    \label{fig:expected_value_compare}
    \vspace{-1.5em}
\end{figure}

\begin{figure}[ht]
    \centering
    \includegraphics[width=0.8\linewidth]{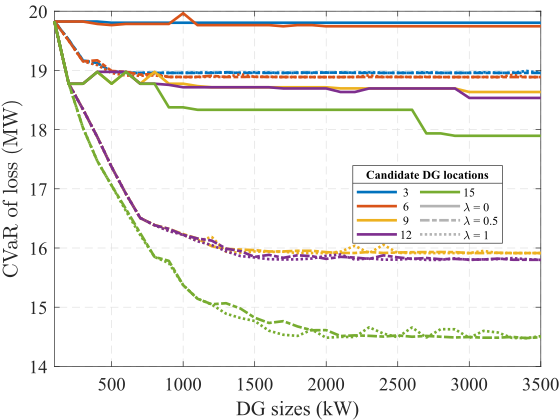}
    \caption{Variation of $CVaR_\alpha$ of prioritized load loss with respect to the number of candidate DG locations, budget allocation, and risk preference.}
    \label{fig:cvar_value_compare}
    \vspace{-1.5em}
\end{figure}

\begin{figure}[ht]
    \centering
    \includegraphics[width=0.8\linewidth]{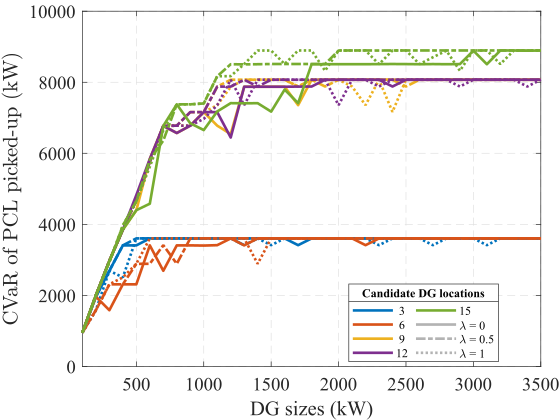}
    \caption{Variation of $CVaR_\alpha$ of prioritized load picked up with respect to the number of candidate DG locations, budget allocation, and risk preference.}
    \label{fig:PCL_cvar_value_compare}
\end{figure}

To analyze the effect of budget allocation, candidate locations, and risk preference, further simulations are conducted. Figs.~\ref{fig:expected_value_compare} - ~\ref{fig:PCL_cvar_value_compare} show the variation of expected value, $CVaR_\alpha$, and $CVaR$ of prioritized critical load (PCL) picked up respectively for $n=49$. For each of the cases, 5 different sets of DG candidate locations are selected and the simulations are conducted for $\sum \beta_i^{DG} \in [100, 3500]$ kW. The expected value of prioritized load loss, as shown in Fig.~\ref{fig:expected_value_compare}, decreases gradually with respect to budget and the number of candidate locations. However, for a fixed number of candidate locations, the expected value of prioritized load loss does not change significantly with a change in risk preference. The $CVaR_\alpha$, however, changes significantly when the number of candidate locations is increased, see Fig.~\ref{fig:cvar_value_compare}. Furthermore, the change is also more pronounced with an increase in risk aversion for a higher number of candidate locations as compared to cases when the number of candidate locations is less; 3 or 6. The result is consistent for the $CVaR$ of PCL pick-up. This means that more prioritized critical loads are picked up during HILP events realization when there are more candidate locations for DGs. These results are consistent with our former conclusion that it is desired to have a smaller number of DGs in different locations than to have fewer DGs with larger capacities for HILP events. However, the results also show that all of these different objectives saturate at some point after which increasing the budget to increase the DG size has no improvement over the loss minimization objective. This is an important conclusion to decide proper budget allocation in any resilience-oriented planning problem.

\vspace{-0.5em}

\section{Conclusion}\label{sec:conclusions}
This paper presents the sensitivity of a resilience-oriented active distribution system planning framework towards several parameters. The analyses were mainly conducted for a different number of scenarios, risk preferences, and budget allocation. The results conclude that there is a trade-off in solution quality vs solve time when varying the number of scenarios. Hence, it is desired to select a number of scenarios that provides considerable planning solutions with reasonable solve time. Furthermore, the overall system operators' budget and the number of candidate DG locations also affect the overall solution. It is desirable to include DGs with smaller sizes in multiple locations when the overall budget is limited. Additionally, at some point, the solution quality does not improve further when more budget is spent on increasing the size of the DG at the same location. As observed in the results, such sensitivity analyses on resilience-oriented planning would provide additional information to system operators when allocating the planning budget.

\bibliographystyle{IEEEtran}
\bibliography{ref.bib}

\end{document}